\documentclass[12pt,epsf]{article}
\usepackage{graphicx,amsmath,amssymb}
\begin{document}

\def\a{\alpha}
\def\b{\beta}
\def\c{\varepsilon}
\def\d{\delta}
\def\e{\epsilon}
\def\f{\phi}
\def\g{\gamma}
\def\h{\theta}
\def\k{\kappa}
\def\l{\lambda}
\def\m{\mu}
\def\n{\nu}
\def\p{\psi}
\def\q{\partial}
\def\r{\rho}
\def\s{\sigma}
\def\t{\tau}
\def\u{\upsilon}
\def\v{\varphi}
\def\w{\omega}
\def\x{\xi}
\def\y{\eta}
\def\z{\zeta}
\def\D{\Delta}
\def\G{\Gamma}
\def\H{\Theta}
\def\L{\Lambda}
\def\F{\Phi}
\def\P{\Psi}
\def\S{\Sigma}

\def\o{\over}
\def\beq{\begin{eqnarray}}
\def\eeq{\end{eqnarray}}
\newcommand{\gsim}{ \mathop{}_{\textstyle \sim}^{\textstyle >} }
\newcommand{\lsim}{ \mathop{}_{\textstyle \sim}^{\textstyle <} }
\newcommand{\vev}[1]{ \left\langle {#1} \right\rangle }
\newcommand{\bra}[1]{ \langle {#1} | }
\newcommand{\ket}[1]{ | {#1} \rangle }
\newcommand{\EV}{ {\rm eV} }
\newcommand{\KEV}{ {\rm keV} }
\newcommand{\MEV}{ {\rm MeV} }
\newcommand{\GEV}{ {\rm GeV} }
\newcommand{\TEV}{ {\rm TeV} }
\def\diag{\mathop{\rm diag}\nolimits}
\def\Spin{\mathop{\rm Spin}}
\def\SO{\mathop{\rm SO}}
\def\O{\mathop{\rm O}}
\def\SU{\mathop{\rm SU}}
\def\U{\mathop{\rm U}}
\def\Sp{\mathop{\rm Sp}}
\def\SL{\mathop{\rm SL}}
\def\tr{\mathop{\rm tr}}

\def\IJMP{Int.~J.~Mod.~Phys. }
\def\MPL{Mod.~Phys.~Lett. }
\def\NP{Nucl.~Phys. }
\def\PL{Phys.~Lett. }
\def\PR{Phys.~Rev. }
\def\PRL{Phys.~Rev.~Lett. }
\def\PTP{Prog.~Theor.~Phys. }
\def\ZP{Z.~Phys. }

\newcommand{\bea}{\begin{eqnarray}}   
\newcommand{\eea}{\end{eqnarray}}
\newcommand{\bear}{\begin{array}}  
\newcommand {\eear}{\end{array}}
\newcommand{\bef}{\begin{figure}}  
\newcommand {\eef}{\end{figure}}
\newcommand{\bec}{\begin{center}}  
\newcommand {\eec}{\end{center}}
\newcommand{\non}{\nonumber}  
\newcommand {\eqn}[1]{\beq {#1}\eeq}
\newcommand{\la}{\left\langle}  
\newcommand{\ra}{\right\rangle}
\newcommand{\ds}{\displaystyle}
\def\SEC#1{Sec.~\ref{#1}}
\def\FIG#1{Fig.~\ref{#1}}
\def\EQ#1{Eq.~(\ref{#1})}
\def\EQS#1{Eqs.~(\ref{#1})}
\def\GEV#1{10^{#1}{\rm\,GeV}}
\def\MEV#1{10^{#1}{\rm\,MeV}}
\def\KEV#1{10^{#1}{\rm\,keV}}
\def\lrf#1#2{ \left(\frac{#1}{#2}\right)}
\def\lrfp#1#2#3{ \left(\frac{#1}{#2} \right)^{#3}}


\baselineskip 0.7cm

\begin{titlepage}

\begin{flushright}
IPMU 11-0020\\
TU-878\\
KEK-TH 1440
\end{flushright}

\vskip 1.35cm
\begin{center}
{\large \bf Number-Theory Dark Matter
}
\vskip 1.2cm
Kazunori Nakayama$^a$,
Fuminobu Takahashi$^{b,c}$
and 
Tsutomu T. Yanagida$^{c,d}$

\vskip 0.4cm

{ \it $^a$Theory Center, KEK, 1-1 Oho, Tsukuba, Ibaraki 305-0801, Japan}\\
{\it $^b$Department of Physics, Tohoku University, Sendai 980-8578, Japan}\\
{\it $^c$Institute for the Physics and Mathematics of the Universe,
University of Tokyo, Kashiwa 277-8568, Japan}\\
{\it $^d$Department of Physics, University of Tokyo, Tokyo 113-0033, Japan}

\vskip 1.5cm

\abstract{ We propose that the stability of dark matter is ensured by
  a discrete subgroup of the U(1)$_{\rm B-L}$ gauge symmetry, $Z_{2}({\rm B-L })$. 
   We introduce a set of chiral fermions charged under the
  U(1)$_{\rm B-L}$ in addition to the right-handed neutrinos, and
  require the anomaly-cancellation conditions associated with the
  U(1)$_{\rm B-L}$ gauge symmetry. We find that the possible number of
  fermions and their charges are tightly constrained, and that
  non-trivial solutions appear when at least five additional chiral
  fermions are introduced. The Fermat theorem in the number theory
  plays an important role in this argument.  Focusing on one of the
  solutions, we show that there is indeed a good candidate for dark
  matter, whose stability is guaranteed by $Z_{2}({\rm B-L })$.  }
\end{center}
\end{titlepage}

\setcounter{page}{2}

The presence of dark matter (DM) in the Universe is clear
observational evidence for the need of physics beyond the standard
model (SM)\footnote{ One exception is a primordial black
  hole~\cite{Frampton:2010sw}, which is the only candidate for DM in
  the SM.  }. The DM abundance has been measured with a very high
accuracy as~\cite{Komatsu:2010fb}
\beq
\Omega_{\rm DM} h^2 \;=\; 0.1109 \pm 0.0056,
\eeq
where $h$ is the present Hubble parameter in units of
$100$\,km/s/Mpc. If the DM is made of as-yet-undiscovered particles,
they must be electrically neutral and long-lived.  In particular, their lifetime must be much longer 
than the present age of the Universe, and the longevity may be ensured by a symmetry such as a
$Z_2$ symmetry\footnote{ Another possibility is a composite dark
  matter scenario~\cite{Hamaguchi:2007rb}.  }.  There are, however, no
exact global continuous or discrete symmetries, according to the
argument on the quantum gravity by Banks and
Seiberg~\cite{Banks:2010zn}.  Therefore, it may well be that the
symmetry is an unbroken subgroup of a gauge symmetry.

The U(1)$_{\rm B-L}$ gauge symmetry is a very attractive symmetry
beyond the SM, since it predicts three families of right-handed
neutrinos. The observed small masses of light neutrinos are naturally
explained by the seesaw mechanism~\cite{seesaw}, provided that the B-L
is broken at a very high energy such as the grand unified theory (GUT)
scale. Interestingly, a $Z_2$ subgroup of the U(1)$_{\rm B-L}$ gauge
symmetry remains unbroken in the low energy, if the U(1)$_{\rm B-L}$
is spontaneously broken by a Higgs field of charge $2$, coupled to the
right-handed neutrinos to generate the large Majorana masses.  Here we
define the normalization of the B-L charge so that the right-handed
neutrinos carry the charge $-1$.  In this letter we consider a
possibility that the $Z_{2}({\rm B-L })$ is responsible for the stability
of DM.

The anomaly-free conditions associated with the U(1)$_{\rm B-L}$
symmetry are satisfied for the SM fermions plus the three right-handed
neutrinos, which however contain no DM candidate\footnote{ If one of
  the right-handed neutrinos is extremely light, it can account for
  DM.  In fact, it is possible to split the mass scales of the
  right-handed neutrinos while keeping the success of the seesaw
  mechanism in the the extra dimensional
  framework~\cite{Kusenko:2010ik}.  }.  Accordingly, we need to
introduce an extra particle, which must be electrically neutral and
(quasi-)stable to be a good candidate for DM.  In this letter we
consider a set of SM-gauge-singlet chiral fermions charged under the
U(1)$_{\rm B-L}$, some of which are to be the DM. One important
constraint on those fermions is that they must satisfy the
anomaly-free conditions associated with the U(1)$_{\rm B-L}$, which
severely limit the possible number of fermions as well as their
(relative) B-L charges. We assume the B-L charges to be rational,
since otherwise we would have an exact global symmetry in
contradiction with the quantum
gravity~\cite{Banks:2010zn}\footnote{The B-L corresponds to the
  fiveness U(1) in the SU(5) GUT. Notice that all the SU(5)-invariant
  operators of quarks and leptons (${\bf 5^*, 10}$) carry integer
  charges of the fiveness U(1).}.  In particular, by observing that
all the SM fermions and the right-handed neutrinos have odd B-L
charges, the lightest new fermion with even B-L charge will be stable and
therefore a candidate for DM.\footnote{ This applies to the case that
  the B-L charges of the additional new fermions are integer.}  Also, if
the B-L charge is relatively large, the fermion mass tends to be light
enough to be produced in the early Universe.

In this letter, we show that one of the ``minimal sets'' of new chiral
fermions contains those with large B-L charges.  Interestingly, one of
them turns out to be a good candidate for the DM in the Universe. Here
we define the minimal set of fermions such that it contains no two
different chiral fermions which have the same B-L charge: i.e. the
number of the $Q_i$ charged chiral fermions $\psi_i$ is one or
zero. Thus, this is one plausible example in which the stability of DM
is ensured by a discrete subgroup of a local symmetry.\footnote{
Other models can be found in Ref.~\cite{Batell:2010bp}, 
in which the stability  of DM is ensured by a discrete symmetry, a remnant of
hidden U(1) gauge symmetry. 
}

\vspace{5mm}

Let us first see how the anomaly-free conditions limit the possible
number and charges of the additional fermions.\footnote{
Mathematical analysis for finding a set of chiral fermions satisfying the anomaly-free
conditions was given in Ref.~\cite{Batra:2005rh}.}  There are two
anomaly-free conditions: one is from $[U(1)_{\rm B-L}]^3$ anomaly and
the other from the gravitational $[U(1)_{\rm B-L}]\times[graviton]^2$
anomaly. They are given by
\begin{equation}
\sum_{i=1}^{n}Q_i^3 =0,  \label{cubic}
\end{equation}
\begin{equation}
\sum_{i=1}^n Q_i =0,              \label{sum}
\end{equation}
where we assume $Q_i \ne 0$, and $n$ is the number of newly introduced
fermions, $\psi_i~(i=1$-$n)$.  We define all the new chiral fermions
to be right-handed fermions without loss of generality.  According to
the argument of Banks and Seiberg~\cite{Banks:2010zn}, $Q_i$ must be a
rational number.  By multiplying a non-zero integer $a$ with the both
sides of Eqs.~(\ref{cubic}) and (\ref{sum}), therefore, we can always
rewrite them in the following form,
\begin{equation}
	\sum_{i=1}^{n} (Z_i)^3 =0,      \label{cubic2}
\end{equation}
\begin{equation}
	\sum_{i=1}^{n} Z_i =0,              \label{sum2}
\end{equation}
where $\{Z_1,\cdots Z_{n} \} = \{aQ_1,\cdots,a Q_n\}$ are integers.
For convenience, we rearrange $\{Z_i\}$ in ascending order: $Z_1 < Z_2
< \cdots < Z_{n}$.  Let us seek solutions of Eqs.~(\ref{cubic2}) and
(\ref{sum2}) with $n>1$.  The solutions of Eqs.~(\ref{cubic}) and
(\ref{sum}) can be obtained by multiplying the integer solutions by
some rational number.

In the case of two new fermions ($n=2$),
one can easily find that the solutions of Eqs.~(\ref{cubic2}) and (\ref{sum2}) are given by $Z_1=-Z_2$.
In this case we can construct a mass term,
\begin{equation}
	\mathcal L = \frac{1}{2} m \psi_1 \psi_2  + {\rm h.c.}.
\end{equation}
The mass parameter $m$ is not constrained by any symmetry and hence
can be as large as the Planck scale.  This does not lead to
interesting phenomenology within our framework, and hence we do not
consider the case of $n=2$. In the following we will exclude
vector-like fermions from the solutions.

In the case of three new fermions ($n=3$), we can easily see that
there is no solution to Eq.~(\ref{cubic2}), by noting the famous
Fermat theorem in the number theory~\cite{Fermat}.\footnote{
	Euler discovered a proof of the Fermat theorem in the special case of the cubic 
	equation (i.e. Eq.~(\ref{cubic2}) with $n=3$)~\cite{Euler},
	which however contained an important missing step. The complete proof was given by Kausler~\cite{Kausler} and many
	others.} 
	 This fact requires
at least four new fermions to cancel the anomalies. Thus we already
suspect that such a solution, if it exists, may contain chiral
fermions of large B-L charges.

In the case of four new fermions ($n=4$), it is easy to prove there is
no phenomenologically interesting solution to the both conditions,
(\ref{cubic2}) and (\ref{sum2}).  If we erase $Z_4$ from the two
equations, we find
\begin{equation}
	(Z_1+Z_2)(Z_2+Z_3)(Z_3+Z_1)\;=\;0.
\end{equation}
Thus, two of $\psi_1$, $\psi_2$ and $\psi_3$ must have the charges of
opposite sign and equal magnitude, and so do the other fermion and
$\psi_4$.  The solution is therefore given by two pairs of vector-like
fermions, in which we are not interested, as in the case of $n=2$.

Thus we are led to consider the case of at least five new chiral
fermions ($n \geq 5$).  Introducing five additional fermions ($n=5$)
is the minimal non-trivial extension.  The independent solutions to
Eqs.~(\ref{cubic2}) and (\ref{sum2}) with $n=5$ and max$\{ |Z_i| \}
\leq 25$ are given in Table~\ref{sol2}.  We find that even this
minimal extension leads to relatively large B-L charges.  We show in
the following that the minimal charge solution in Table~\ref{sol2}
contains a good candidate for the DM.

\begin{table}[t!]
  \begin{center}
    \begin{tabular}{ | c | c | c | c | c |}
      \hline 
         $Z_1$  & $Z_2$ & $Z_3$ & $Z_4$ & $Z_5$ \\
       \hline 
          $-9$ & $-5$ & $-1$ & $7$ & $8$ \\ \hline
          $-9$ & $-7$ & $2$ & $4$ & $10$ \\ \hline
          $-18$ & $-17$ & $1$ & $14$ & $20$ \\ \hline
          $-21$ & $-12$ & $5$ & $6$ & $22$ \\ \hline
          $-25$ & $-8$ & $-7$ & $18$ & $22$ \\ \hline        	
    \end{tabular}
    \caption{ 
	Independent  solutions to Eqs.~(\ref{cubic2}) and (\ref{sum2}) for max$\{ |Z_i| \} \leq 25$ for $n=5$.
     }
    \label{sol2}
  \end{center}
\end{table}

The seesaw mechanism~\cite{seesaw} for neutrino mass generation
suggests the Majorana mass of the (heaviest) right-handed neutrino at
about the GUT scale.  For this purpose, we introduce a Higgs field
$\Phi$ with the B-L charge $2$.  We assume that $\Phi$ develops a
vacuum expectation value (vev), $\la \Phi \ra = v_{\rm B-L}$, where
$v_{\rm B-L}$ represents the B-L breaking scale.  The three
right-handed neutrinos, $N_1$, $N_2$ and $N_3$, acquire a mass from
\beq
\frac{1}{2} \kappa_3 \Phi N_3 N_3 +\frac{1}{2} \kappa_2 \Phi N_2 N_2 + \frac{1}{2} \kappa_1 \Phi N_1 N_1,
\eeq
where $\kappa_{1,2,3}$ are coupling constants with $|\kappa_3|\geq
|\kappa_2| \geq |\kappa_1|$, and we have adopted a basis such that the
right-handed neutrino mass matrix is diagonalized.  The breaking scale
can be inferred from the neutrino oscillation data, assuming that the
couplings of the heaviest right-handed neutrino are of order unity.
Then the B-L breaking scale $v_{\rm B-L}$ is estimated to be about
$\GEV{15}$.  As noted before, there is unbroken $Z_{2}({\rm B-L })$,
as long as the U(1)$_{\rm B-L}$ symmetry is broken only by the vev of
$\Phi$. Since $Z_{2}({\rm B-L })$ is a subgroup of the gauge symmetry,
it can be an exact symmetry which ensures the stability of DM.

First, let us focus on the integer solutions to Eqs.~(\ref{cubic}) and
(\ref{sum}), which are obtained by identifying
$Z_i$ with $\pm Q_i$, i.e. by setting $a=\pm 1$.  To be explicit we
consider one of the integer solutions,
\beq
(Q_1,Q_2,Q_3,Q_4,Q_5) = (-9, -5, -1, 7, 8),
\label{first_sol}
\eeq
which corresponds to the first solution in Table~\ref{sol2} with $a =
1$.  In this example, the chiral fermion with a charge $8$, $\psi_5$,
is stable because this is the only chiral fermion with an even B-L
charge and it has no mixings with the other fermions.  In other words,
the stability of $\psi_5$ is guaranteed by the $Z_{2}({\rm B-L })$.
 Therefore the fermion $\psi_5$ is a prime candidate for
DM in this solution.

The $\psi_5$ acquires a mass from the following non-renormalizable
operator,
\beq
\frac{\Phi^{*8}}{M^7} \psi_5 \psi_5,
\eeq
where $M$ is a cut-off scale. Then the mass of $\psi_5$ is 
given by
\beq
m_{\psi_5} \;\approx 10{\rm\,keV} \lrfp{v_{\rm B-L}}{3\times \GEV{15}}{8}    \label{mpsi}
\eeq
where we have set the cut-off scale $M$ to be the Planck mass, $M_P
\approx 2.4 \times \GEV{18}$. In the following we will take $v_{\rm
  B-L} = 3 \times \GEV{15}$ and $M=M_P$ as reference values unless
otherwise stated. The mass of $\psi_5$ is sensitive to the B-L
breaking scale, but it can be heavy enough to be consistent with the
Lyman-alpha data~\cite{Boyarsky:2008xj}.

In order to account for the DM, $\psi_5$ must have the correct
abundance.  The main production process is pair production of $\psi_5$
from the SM fermions in plasma through the s-channel exchange of the
B-L gauge boson~\cite{Kusenko:2010ik}.  The production is most
efficient at reheating.  The $\psi_5$ number to entropy ratio can be
roughly estimated as
\bea
Y_{\psi_5}\; \equiv\; \frac{n_{\psi_5}}{s}  &\sim& \left.\frac{\la \sigma v \ra n_f^2/H}{\frac{2 \pi^2}{45}g_* T^3}\right|_{T=T_R} \non\\
			&\sim &4 \times 10^{-5} \lrfp{g_*}{10^2}{-\frac{3}{2}} \lrfp{Q_5}{8}{2}
			 \lrfp{v_{\rm B-L}}{3\times \GEV{15}}{-4} \lrfp{T_R}{4 \times \GEV{13}}{3},   \label{Y}
\eea
where $H$ is the Hubble parameter, $g_*$ counts the relativistic
degrees of freedom at the reheating, $\la \sigma v \ra \sim T^2/v_{\rm
  B-L}^4$ is the production cross section, $n_f \sim T^3$ is the
number density of the SM fermions in plasma, 
$T_R$ is the reheating temperature and the first equality is
evaluated at the reheating.  The numerical solution of the Boltzmann
equation gives a consistent result~\cite{Khalil:2008kp}.  The DM
abundance is given by
\beq
\Omega_{\psi_5}h^2 \;\approx\;0.1  \left(\frac{m_{\psi_5}}{10\, {\rm keV}}\right)  \left( \frac{Y_{\psi_5}}{4 \times 10^{-5}} \right).
\label{Omegadm}
\eeq
Thus, the reheating temperature as high as $O(10^{13})$\,GeV is needed to
account for  the DM density by this production process.
The thermal leptogenesis~\cite{Fukugita:1986hr} works with such a high temperature.

Let us comment on the properties of the other fermions, $\psi_1,
\psi_2, \psi_3$, and $\psi_4$, which have charges of $-9,-5,-1$ and
$7$, respectively.  We fist note that $\psi_3 (-1)$ will have a
Majorana mass of order $v_{\rm B-L}$ similarly to the right-handed
neutrinos.  We can also see that $\psi_4$ and one linear combination
of $\psi_1$ and $\psi_2$ acquire a mass of order $v_{\rm B-L}$ from
the mass terms, $\Phi \psi_1 \psi_4$ and $\Phi^* \psi_2 \psi_4$;
these heavy modes do not significantly contribute to the light neutrino
masses, and so they can be safely integrated out.  Also they do not
have any cosmological effects, since they are not produced in the
early Universe if the reheating temperature is smaller than their
masses.  On the other hand, the other orthogonal combination of
$\psi_1$ and $\psi_2$, denoted by $\psi_{\ell}$ in the following,
remains light.  The $\psi_{\ell}$ obtains a mass of order $v_{\rm B-L}^5/M^4 \sim  7$\,TeV
both from the Majorana mass term and from  the mixing with the right-handed neutrinos.
The $\psi_{\ell}$ quickly decays into a lepton and a Higgs boson before the big bang
nucleosynthesis, and therefore has no drastic effects on cosmology.\footnote{
The lepton asymmetry produced by the right-handed neutrino decay is not washed
out if the mixing of $\psi_\ell$ with the right-handed neutrinos is suppressed
by $O(0.1)$.
}

Let us comment on a possible problem with other solutions. For instance, we consider 
the first solution in Table~\ref{sol2} with $a = -1$. Similarly, we would then have a light fermion $\psi_{\ell}^\prime$.
Problem is that $\psi_{\ell}^\prime$ would generally have larger mixings with
the right-handed neutrinos, which spoils the seesaw formula. Thus, for such solutions,
dangerous mixings between $\psi_\ell^\prime$ and the right-handed neutrinos must be suppressed
by introducing an approximate matter parity or extra dimensional set-up,
in order to retain the seesaw mechanism. 

Next let us consider a rational number solution to Eqs.~(\ref{cubic})
and (\ref{sum}).  One advantage of rational number solutions is that
possible dangerous mixings with the right-handed neutrinos can be suppressed. 
 On the other hand, the stability of the lightest $Z_{2}({\rm B-L })$
  even fermion is not necessarily ensured because of the
presence of fermions of fractional charge. Let us consider one
example.  Multiplying the previous solution with $1/2$, we obtain
\beq
(Q_1,Q_2,Q_3,Q_4,Q_5) \;=\; \left(-\frac{9}{2}, -\frac{5}{2}, -\frac{1}{2},\, \frac{7}{2}, \,4 \right).
\label{frac}
\eeq
As before, $\psi_5$ is a candidate for DM and it has a mass of $v_{\rm
  B-L}^4/M^3 \sim 6\times 10^6{\rm\, GeV}$ for the same reference
values of $v_{\rm B-L}$ and $M$ as those in the previous case.  The
correct abundance is realized if the reheating temperature is $T_R
\sim 8\times 10^9$\,GeV.\footnote{ Since the number-to-entropy ratio
  is proportional to $v_{\rm B-L}^{-4}$ (see (\ref{Y})), the DM relic
  density (\ref{Omegadm}) does not depend on $v_{\rm B-L}$.  Thus the
  $\psi_5$ mass can be lighter if we choose a smaller value of $v_{\rm
    B-L}$.  } The thermal leptogenesis works for such high $T_R$.  The
other additional fermions have fractional charges and we can see that
they remain massless and their contributions to the extra radiation
will be negligibly small.  Moreover, since they have fractional
charges, they do not mix with the SM fermion and the right-handed
neutrinos.   We note here that
$\psi_5$ is not absolutely stable, because of the presence of fermions
of fractional charge.  In fact it mainly decays into $\psi_2$, $\psi_3$, the
SM lepton and the Higgs boson with a lifetime more than twenty orders
of magnitude longer than the present age of the Universe, for the
reference values of $v_{\rm B-L}$ and $M$.  The lifetime is
proportional to high powers of $v_{\rm B-L}$, and so, the decay could
contribute to the high energy cosmic-ray spectrum for a slightly large
value of $v_{\rm B-L}$.  Note that another solution can be obtained by
multiplying (\ref{frac}) with an odd integer, which however leads to
too light DM mass.

Finally we briefly discuss other solutions.  Let us focus on the
integer solution of $(Q_1,Q_2, Q_3, Q_4, Q_5)=(-18,-17,1,14,20)$.
There are three $Z_{2}({\rm B-L })$ even fermions, one of which may be a
candidate for DM.  For a similar reason discussed above, one linear
combination of $\psi_4$ and $\psi_5$ remains light.  Its mass is
however of the order of $\sim v_{\rm B-L}^{14}/M^{13}$, and it is
extremely light.  The $Z_{2}({\rm B-L })$ odd fermion, $\psi_2$, is also
extremely light and it can be regarded as a stable particle in a
cosmological time scale.  Thus the light and stable particles in this
solution are too light to be DM.  Similar consideration leads us to
conclude that additional fermions tend to be extremely light and DM
candidates do not likely exist as the U(1)$_{\rm B-L}$ charges become
larger.  On the other hand, these light stable particles may
constitute a part of extra radiation in the Universe.  If the number
of such light fermions is large enough, they may be able to explain
the extra radiation of the Universe indicated by recent
observations~\cite{Komatsu:2010fb,Izotov:2010ca,Nakayama:2010vs,Krauss:2010xg}.

\vspace{5mm}

Let us comment on other implications of our scenario. So far we have implicitly assumed that
the U(1)$_{\rm B-L}$ symmetry is spontaneously broken during inflation and it is not restored after inflation.
However, it is possible that the U(1)$_{\rm B-L}$ gets spontaneously broken after inflation, if the
mass of $\Phi$ is light enough. Then the cosmic strings will be 
formed at the phase transition, which can be probed by the CMB measurement such as the Planck satellite.
In general the DM abundance is modified in this case, but the estimate (\ref{Y}) remains valid
as long as the phase transition occurs at $T \gtrsim \GEV{14}$. Another implication of the integer
solution (\ref{first_sol}) is that the DM mass is in the keV range so that
it can affect the structure formation at small scales. We note that the lifetime of DM depends on
the assumption that U(1)$_{\rm B-L}$ is spontaneously broken only by $\Phi$.
  If $Z_{2}({\rm B-L })$ is also broken by another Higgs field of a B-L odd charge, the
DM decays into the SM particle, which may produce observable signature
in the indirect DM search.  For instance, the DM decay produces an
X-ray photon in the case of (\ref{first_sol}), while high-energy
gamma-ray, anti-protons, and positrons are produced in the case of (\ref{frac}).  
In fact,  even without the additional Higgs, 
the DM is not absolutely stable in case of  (\ref{frac}), 
and the decay can contribute the cosmic rays if the B-L breaking
scale is several times larger than our reference value.
On the other hand, no signals are expected in the direct DM search. 
We have seen that a high reheating temperature
is needed in order to obtain a correct DM abundance~(\ref{Y}) in the
case of (\ref{first_sol}). Since the inflaton mass must be larger than
the reheating temperature~\cite{Kolb:2003ke}, the chaotic
inflation~\cite{Linde:1983gd} is one of the possibilities.  This may
be confirmed by the observation of the B-mode in the cosmic microwave
background polarization.

\vspace{5mm} In this letter we have considered a possibility that the
longevity of DM is guaranteed by the $Z_{2}({\rm B-L })$.  To this end
we have introduced $n$ SM-gauge-singlet chiral fermions charged under
the U(1)$_{\rm B-L}$, in addition to the three right-handed
neutrinos. We have shown that more than four additional chiral
fermions are required for the cancellation of the anomalies associated
with the U(1)$_{\rm B-L}$.  In the case of $n=2$, two fermions must
have charges of opposite sign and same magnitude, and they can be
integrated out if their mass is as heavy as the Planck scale.
Importantly, the Fermat theorem excludes the case of $n=3$, which have
forced us to consider $n > 3$.  In the case of $n=4$, the four
fermions are divided into two pairs of vector-like fermions, which
have charges of opposite sign and same magnitude, and this case was
not of our interest.  Thus, we are led to consider more than four
additional chiral fermions.  Interestingly, one of the minimal set of
the integer B-L charges satisfying the anomaly-free conditions contains
a good candidate for DM.  The DM abundance can be explained if the
reheating  temperature is as high as $O(10^{13})$\,GeV or $O(10^9)$\,GeV,
depending on whether the charges are integer or fractional. Both cases are
consistent with the thermal leptogenesis scenario.

\section*{Acknowledgment}

We would like to thank Satoshi Kondo for useful discussions and
particularly for his interdisciplinary colloquium on Number Theory at IPMU,
where the basic idea of the present letter occurred to us.  This work
was supported by the Grant-in-Aid for Scientific Research on
Innovative Areas (No. 21111006) [KN and FT], Scientific Research (A)
(No. 22244030 [FT] and 22244021 [TTY]), and JSPS Grant-in-Aid for
Young Scientists (B) (No. 21740160) [FT].  This work was also
supported by World Premier International Center Initiative (WPI
Program), MEXT, Japan.

\end{document}